\documentclass[reprint,aps,prl,twocolumn,superscriptaddress,showpacs,floatfix,preprintnumbers]{revtex4-1}
\usepackage{color}
\usepackage{mathtools}

\usepackage{ifpdf}
\usepackage{cancel}

\ifpdf
  \usepackage[pdftex]{graphicx}
  \DeclareGraphicsExtensions{.pdf}
  \usepackage[hyperindex=true]{hyperref}
\else
  \usepackage{graphicx}
  \DeclareGraphicsExtensions{.ps,.eps}
  
  \usepackage[a4paper,dvipdfm,hyperindex=true]{hyperref}
\fi

\definecolor{linkcol}{rgb}{0.2,0.2,0.6}

\hypersetup{bookmarksopen=true,
pdfmenubar=true, %menubar shown
pdfhighlight=/O, %effect of clicking on a link
colorlinks=true, %couleurs sur les liens hypertextes
pdfpagemode=None, %aucun mode de page
pdffitwindow=true, %pages ouvertes entierement dans toute la fenetre
linkcolor=linkcol, %couleur des liens hypertextes internes
citecolor=linkcol, %couleur des liens pour les citations
urlcolor=linkcol %couleur des liens pour les url
}

\def\Qb{\ensuremath{|\mathbf{Q}|}}
\def\SCTO{Sr$_2$CuTeO$_6$}
\def\SCWO{Sr$_2$CuWO$_6$}
\def\SCTWO{Sr$_2$CuTe$_x$W$_{1-x}$O$_6$}
\def\SCTWOf{Sr$_2$CuTe$_{0.5}$W$_{0.5}$O$_6$}

\newcommand{\RNum}[1]{\uppercase\expandafter{\romannumeral #1\relax}}

\begin{document}

%\preprint{DRAFT 9}

%\title{Origin of magnetic disorder in Sr$_2$CuTe$_{x}$W$_{1-x}$O$_6$}
\title{Exchange Interactions Mediated by Non-Magnetic Cations in Double Perovskites}

\author{Vamshi M. Katukuri}
\email{vamshi.katukuri@epfl.ch}
\affiliation{Chair of Computational Condensed Matter Physics, Institute of Physics, \'{E}cole Polytechnique F\'{e}d\'{e}rale de Lausanne (EPFL), CH-1015 Lausanne, Switzerland}
\author{P. Babkevich}
\email{peter.babkevich@epfl.ch}
\affiliation{Laboratory for Quantum Magnetism, Institute of Physics, \'{E}cole Polytechnique F\'{e}d\'{e}rale de Lausanne (EPFL), CH-1015 Lausanne, Switzerland}
\author{O. Mustonen}
\affiliation{Department of Chemistry and Materials Science, Aalto University, FI-00076 Espoo, Finland}
\affiliation{Department of Material Science and Engineering, University of Sheffield, Mappin Street, Sheffield, S1 3JD, United Kingdom}
\author{H. C. Walker}
\affiliation{ISIS Neutron and Muon Source, Rutherford Appleton Laboratory, Chilton, Didcot, OX11 OQX, United Kingdom}
\author{B. F{\aa}k}
\affiliation{Institut Laue-Langevin, CS 20156, F-38042 Grenoble Cedex 9, France}
\author{S. Vasala}
\affiliation{Institut f\"ur Materialwissenschaft, Fachgebiet Materialdesign durch Synthese, Technische Universit\"{u}t Darmstadt, Alarich-Weiss-Stra\ss e 2, 64287 Darmstadt, Germany}
\author{M. Karppinen}
\affiliation{Department of Chemistry and Materials Science, Aalto University, FI-00076 Espoo, Finland}
\author{H. M. R\o nnow}
\affiliation{Laboratory for Quantum Magnetism, Institute of Physics, \'{E}cole Polytechnique F\'{e}d\'{e}rale de Lausanne (EPFL), CH-1015 Lausanne, Switzerland}
\author{O. V. Yazyev}
\affiliation{Chair of Computational Condensed Matter Physics, Institute of Physics, \'{E}cole Polytechnique F\'{e}d\'{e}rale de Lausanne (EPFL), CH-1015 Lausanne, Switzerland}

\begin{abstract}

Establishing the physical mechanism governing exchange interactions is fundamental for exploring exotic phases such as quantum spin liquids (QSLs) in real materials. 
In this work, we address exchange interactions in \SCTWO, a series of double perovskites that realize a spin-1/2 square lattice and are suggested to harbor a QSL ground state arising from the random distribution of non-magnetic ions. 
Our \textit{ab initio } multi-reference configuration interaction calculations show that replacing Te atoms with W atoms changes the dominant couplings from nearest to next-nearest neighbor due to the crucial role of unoccupied states of the non-magnetic ions in the super-superexchange mechanism. 
Combined with spin-wave theory simulations, our calculated exchange couplings provide an excellent description of the inelastic neutron scattering spectra of the parent compounds, as well as explaining that the magnetic excitations in \SCTWOf\ emerge from bond-disordered exchange couplings. 
Our results demonstrate the crucial  role of the non-magnetic cations in exchange interactions paving the way to further explore QSL phases in bond-disordered materials.
\end{abstract}

%\pacs{}
\date{\today}
\maketitle
In 3d transition metal (TM) oxides, the on-site Coulomb repulsive interactions between the electrons are strong enough to confine them to the TM sites, leading to the formation of localized spin or spin-orbital moments~\cite{khomskii-book}. 
The manner in which these moments couple to each other is primarily governed by the underlying exchange interactions, which may be direct and/or mediated by the intermediate anions or ligands (L), the latter is also referred to as the superexchange.
There are many possible ways these interactions can manifest, resulting in a plethora of magnetically ordered states such as ferromagnetic and different types of antiferromagnetic (AFM) order, magnetic spirals or more exotic topologically protected magnetic textures such as Skyrmions~\cite{khomskii-book,spirals_review,Roessler_nature_2006,Muehlbauer_science_2009}.

Even more fascinating ground states that stem from exchange interactions are those which do not undergo any magnetic ordering even at absolute zero temperature, e.g. spin-liquid states in low-dimensional magnetic systems~\cite{balents-nature-2010}. % such as  one or two dimensions. 
Broken-symmetry valence-bond solids and QSLs where symmetry is conserved are examples of such phases~\cite{anderson-1973, anderson-science-1987, balents-nature-2010}.
In these quantum paramagnetic phases, the long-range magnetic order is typically destroyed by frustrated exchange interactions and quantum fluctuations~\cite{Mila_frustrated_magnetism}. 
%{\cbl 
	In the simplistic and prototypical two-dimensional spin-$ 1/2 $ Heisenberg  square lattice (HSL) model, the ratio of nearest-neighbor (NN) $ J_{1} $ and AF next-nearest neighbor (NNN) $ J_{2} $ exchange interactions of $\sim$0.5 results in magnetic frustration and a QSL ground state~\cite{anderson-science-1987}. 
%}
%For example, magnetic frustration kicks in the most simplistic two-dimensional spin-$ 1/2 $ Heisenberg  square lattice (HSL) model with a QSL groun state when the ratio of nearest-neighbor (NN) $ J_{1} $ and AF next-nearest neighbor $ J_{2} $ exchange interactions is close to 0.5~\cite{anderson-science-1987}.  
 % in the spin-$1/2$ Heisenberg square-lattice (HSL) model with ferro or AF nearest-neighbor (NN) $J_1$ and AF next-nearest neighbor (NNN) $J_2$ exchange interactions, 
 % $J_2/|J_1|$ $\approx$ 0.5 results in magnetic frustration giving rise to a QSL~\cite{anderson-science-1987}. 
   
 %Despite the requirements for a frustrated $J_1$-$J_2$ spin-$1/2$ HSL model are not exigent, there is no experimental realisation of the model yet.
%
The exchange mechanisms in TM compounds, principally the superexchange, are reasonably well understood in the form of the Goodenough-Kanamori-Anderson (GKA) rules~\cite{khomskii-book}. 
The highly successful GKA rules correctly predict the sign of magnetic coupling for the 180$ ^\circ $ and 90$ ^\circ $ TM-L-TM bond angles.  
In double perovskite compounds like \SCTO\ and \SCWO\, the magnetic Cu$^{2+}$ ions are separated by non-magnetic Te$ ^{6+} $ and W$ ^{6+} $ cations, respectively, and the magnetic coupling is a result of the super-superexchange (SSE) mechanism.
As shown in Fig. 1, there are multiple SSE paths -- the NN exchange is via two identical Cu-O-Te/W-O-Cu paths involving four bridging ligands and two non-magnetic cations, with a 90$ ^\circ $ Cu-Te/W-Cu angle. 
 Alternatively,  the second or NNN coupling arises from only one Cu-O-Te/W-O-Cu exchange path (180$ ^\circ $ Cu-Te/W-Cu angle) involving two ligands and a non-magnetic cation. 
An interesting aspect in these compounds is weather the non-magnetic cation participates in the exchange.

In this paper, we address the question -- ``Do the exchange mechanisms in double perovskite compounds depend on the non-magnetic cations and if so, how?''  
We compute the exchange couplings in double perovskite \SCTWO\ with $x=\{0.0, 0.5, 1.0\}$ compounds using {\it ab initio} many-body quantum chemistry (QC) calculations.
We analyze their microscopic provenance by 
 examining the different SSE paths involved and show that the bridging non-magnetic cation plays a pivotal role in the exchange mechanism depending on whether it has an empty or completely filled $d$ manifold~\cite{Zhu_cr_w_te_prl_2014,Zhu_cr_dp_2015,Mustonen2018}.
Further, we decipher the possible physical origin of the features observed in the inelastic neutron scattering (INS) spectrum using spin-wave theory (SWT) while simulating the site disorder phenomena in \SCTWO .
%{\cbl 
	Our study exposes double perovskites with non-magnetic cations as the ideal playground to explore bond-disordered couplings and associated QSL phenomena.
%} 
%In this paper, we investigate how the exchange couplings manifest in 

% Fig. 1 : Crystal structure
\begin{figure}[!t]
	
	\includegraphics[width=0.95\columnwidth]{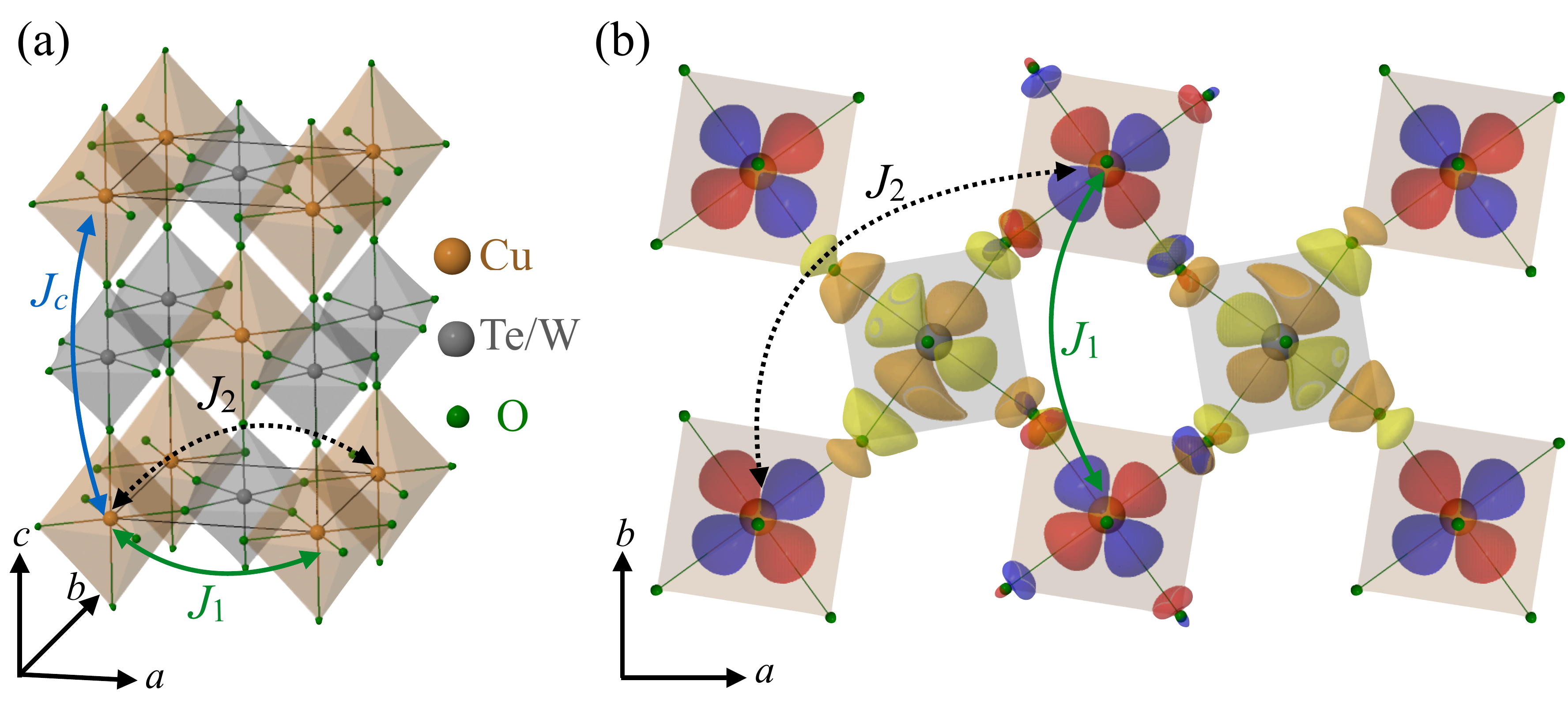}
	\caption{(a) The crystallographic unit cell of Sr$_2$Cu(Te/W)O$_6$. (b) The $ab$-plane view of \SCWO\ showing the in-plane Cu $ 3d_{x^2-y^2} $ (red and blue) and W $ 5d_{x^2-y^2} $ (yellow and orange) orbitals, and the different exchange couplings.}
	% are shown by arrows connecting two Cu$^{2+}$ ions.}
	\label{Fig1}
\end{figure}
The isostructural double perovskite copper oxides \SCTO\ ~\cite{Reinen_SCTO_Struct_ZAAC} and \SCWO\ ~\cite{vasala_SCWO_structure_2012} realise a quasi-two-dimensional spin-$1/2$ HSL antiferromagnet, despite their three-dimensional crystal structures~\cite{koga-jpn-2014,babkevich_prl_2016,walker-prb-2016}.
However, the magnetic order in the two systems is different. 
While a  N\' eel AFM (NAF) ordering is observed in \SCTO\  with large AFM $J_1$ and small $J_2$ of the same sign~\cite{koga-prb-2016,babkevich_prl_2016}, 
a columnar AFM (CAF) order is stabilized in \SCWO\ with small $J_1$ and large $J_2$, both AFM~\cite{walker-prb-2016}.
Interestingly, the reported $J_2/J_1$ ratio in these two compounds differs by two orders of magnitude, 0.03 and 7.92 for \SCTO\ and \SCWO, respectively. 

It has been anticipated that a solid solution with equal quantities of Te and W may result in the ratio $J_2/J_1$ close to 0.5, leading to strong magnetic frustration and possibly producing a spin-liquid ground state~\cite{Mustonen2018,Watanabe_scwto_vbg_prb_2018,Kazuki_randomness_qsl_2018}.
Interestingly, the macroscopic magnetic features of Sr$_2$CuTe$_{x}$W$_{1-x}$O$_6$ for $x=0.5$ show no signs of magnetic ordering, instead indicates its proximity to the highly frustrated $J_2/J_1=0.5$ region.
Furthermore, the specific heat behavior at low temperatures is reminiscent of a gapless QSL state with collective excitations of entangled spins~\cite{Mustonen2018}.
Fascinatingly, the suppression of long range magnetic order is observed in a wide region of $x \approx 0.1-0.6$~\cite{mustonen-prb-2018}.

{\it Exchange couplings from QC calculations}: Table~\ref{Exchg_inter} compares the 
 %NN ($J_1$) and NNN ($J_2$) 
 Heisenberg exchange couplings defined in Fig.~\ref{Fig1} for the end compounds of \SCTWO\, solid solution -- \SCTO\ and \SCWO\, -- obtained from {\it ab initio} multireference difference dedicated configuration interaction (MR-DDCI) calculations~\cite{DDCI_1_MIRALLES1992555,DDCI_2_MIRALLES199333}. 
 Those obtained from INS measurements are also shown in the same table. 
The calculations were done on three different embedded clusters for $J_1$, $J_2$ and $J_c$ (see Fig.~\ref{Fig1}), respectively. 
 For computational details see Ref.~\cite{babkevich_prl_2016} and Supplementary material~\footnote{Supplementary information [URL] contains the computational details.} which includes Refs.~\cite{NOCI_J_hozoi03,ewald,FIGGEN2005227,Peterson2005,Dunning89,Andrae1990,Martin2001,Ross_W_ecp_1990,ANO-S_O_basis,J_ligand_fink94,J_ligand_calzado03,NOCI_J_oosten96,Ir113_bogdanov_12,localization_PM,Molpro12,Davidson_MRCI}.
%{\cbl
	 In contrast to conventional density functional theory (DFT) and correlated calculations in conjunction with dynamical mean field theory (DFT + DMFT), our calculations are parameter free and accurately describe correlations within the cluster of atoms in a systematic manner. 
	 The virtual hopping processes necessary to capture the exchange interactions are well described in this approach and this makes it the only {\it ab initio} method that has sufficient predictive capability for estimating magnetic couplings~\cite{Li2Cu2O2_Js,Ir214_katukuri_12,Ir213_katukuri_13,Ir214_katukuri_14}.
 %}
 %  
  To extract the isotropic exchange couplings, the {\it ab initio} magnetic spectrum of two unpaired electrons in two Cu$^{2+}$ ions is mapped onto that of a two-spin Heisenberg Hamiltonian ${\mathcal H}_{ij}=J_{ij}\mathbf{S}_i\cdot\mathbf{S}_j$.
%{\cred \sout{In the QC computations, a complete active space reference wavefunction with two electrons in the two Cu$^{2+}$ ground state $d_{x^2-y^2}$-type orbitals was first constructed for the singlet and triplet spin multiplicities~\cite{coen_degraf_book}, state-averaged.
		%~\footnote{As for \SCTO\ ~\cite{babkevich_prl_2016}, we find the single unpaired Cu 3$d$ electron in $d_{x^2-y^2}$ orbital.}. % ~\cite{Note2}.
%In the MR-DDCI calculations the electrons in the doubly occupied Cu $3d$ orbitals and the O $2p$ orbitals of the bridging (Te/W)O$_6$ octahedron were correlated.
%To account for size-consistency errors, we adopted Davidson correction scheme~\cite{Davidson_MRCI}. }
%}
All calculations were done using the {\sc molpro} quantum chemistry package~\cite{Molpro12}.

%TABLE I
\begin{table}[!b]
	\caption{
		A comparision of the Heisenberg exchange couplings obtained from {\em ab initio} MR-DDCI calculations (QC) and experimentally using inelastic neutron scattering (INS) for \SCTO~\cite{babkevich_prl_2016} and \SCWO \cite{walker-prb-2016}. The couplings obtained from INS are from fits to SWT with a renormalization factor $Z_c = 1.18$ as the first-order correction~\cite{singh-prb-1989} to calculated magnetic dispersion. All values are given in meV.
	}
	\label{Exchg_inter}
	%\begin{tabular}{ccccc}
	\begin{tabular}{p{1cm}cp{2.10cm}cp{1.5cm}cp{2.10cm}cp{1.5cm}c}
		\hline
		\hline\\[-0.40cm]
		&\multicolumn{2}{l}{ \SCTO}  &\multicolumn{2}{l}{\SCWO} \\
		\hline
		& QC & INS~\cite{babkevich_prl_2016}    &QC & INS~\cite{walker-prb-2016}\\
		\hline\\[-0.25cm]
		$J_1$  & 7.38  & 7.60(3)  &  0.68  & 1.02\\
		$J_2$  & 0.05  & 0.60(3)  &  8.33  & 8.50\\
		$J_c$  & 0.003 & 0.04     &  0.005 & -\\
		\hline
		\hline
	\end{tabular}
\end{table}

We have previously shown~\cite{babkevich_prl_2016} that in \SCTO\ the dominant Heisenberg coupling is the NN AF $J_1$, see columns one and two in Table~\ref{Exchg_inter}. 
The SSE path that gives this large coupling is Cu$^{2+}$-O$^{2-}$-O$^{2-}$-Cu$^{2+}$ along the two bridging TeO$_6$ octahedra and does not include the Te$ ^{6+} $ ions explicitly.
On the other hand, the NNN $J_2$ coupling with 180$^{\circ}$ Cu-Te-Cu angle is significantly smaller and is through the bridging Te atom -- Cu$^{2+}$-O$^{2-}$-Te$^{6+}$-O$^{2-}$-Cu$^{2+}$. 

We performed QC calculations for \SCWO\ to find a strong NNN $J_2$ and a small NN $J_1$ resulting in the ratio $J_2/J_1 \sim 12$.
The coupling along the $c$-axis is estimated to be two orders of magnitude smaller, but nevertheless larger than in \SCTO. 
These results are consistent with the couplings extracted from the INS data~\cite{walker-prb-2016}. 
%\added{
	We note that the magnon-magnon interaction, not included in the linear SWT employed in Ref.~\cite{walker-prb-2016}, could have the same effect as a small NN exchange coupling. 
	Therefore, further corrections to $J_1$ and $J_2$ values extracted from the INS might be necessary to account for this. 
	%While this effect should decrease the magnitude of the couplings, making the QC and experimental numbers rather close, at the same time this decrease is 
	However, we expect this to be small, e.g. in \SCTO, this corresponds to $(J_1,J_2)$ values being renormalized from (7.60,0.60) to (7.18,0.21)~\cite{babkevich_prl_2016}.
%}

Given the qualitative similarity of the crystal structures~\cite{Reinen_SCTO_Struct_ZAAC,vasala_SCWO_structure_2012} as well as the electronic states near the Fermi level~\cite{Yuanhui_DFT_2017}, 
it seems surprising to find the dominant $ab$-plane exchange couplings reversed in the two compounds -- $J_1$ in \SCTO\ and $J_2$ in \SCWO.
It is important to note that the states above (unoccupied) and below (doubly occupied) the Fermi level play an active role in the superexchange process, 
particularly, if these states belong to the ions bridging the two magnetic sites. 
In this respect, there is a considerable difference in the unoccupied manifold near the Fermi level in the two compounds. 
While there is a large density of W 5$d$ unoccupied states in \SCWO\ at 4 eV above the Fermi level~\cite{Yuanhui_DFT_2017}, in \SCTO\ the relatively small density of unoccupied states near the Fermi level consists of Te 5$p$ character~\cite{Yuanhui_DFT_2017}. 
Further, owing to the delocalized nature of W $5d$ orbitals, there is a considerable $dp$-hybridization with the bridging O 2$p$ orbitals leading to appreciable hopping matrix element across the W$^{6+}$ ions. 
In contrast, the Te 5$p$ orbitals are compact and little or zero $pp$-hybridization is expected with O 2$p$ orbitals. Thus, in \SCWO\ the W$^{6+}$ ions actively participate in the superexchange mechanism.  

One might ask ``why $J_1$ is small in \SCWO\ ?", given the arguments brought forward in the previous paragraph. 
To gain more insight into the SSE paths involved in NN and NNN couplings, we have computed $J_1$ and $J_2$ by restraining the virtual hopping processes involving the W$ ^{6+} $ unoccupied orbitals.  
This can be achieved in QC calculations by setting the coefficients of these orbitals to zero. 
Although this is unphysical, it gives direct information about the role of W virtual orbitals in the exchange mechanisms. 

The NN $J_1$ coupling along two 90$^{\circ}$ Cu-W-Cu paths, see Fig.~\ref{Fig1}(b), decreases to 0.49 meV ($\sim$25\% reduction) when the unoccupied orbital coefficients of the two W$ ^{6+} $ ions are eliminated. 
This implies that the contribution to $J_1$ from the configurations involving W virtual orbitals is not the leading one. 
It turns out that the other exchange paths, particularly the Cu-O-O-Cu path, has larger contribution to $J_1$ just as in the case of \SCTO~\cite{babkevich_prl_2016}. %that provide a dominant contribution.
It should be noted that there are three different W $5d$ orbitals that participate in the SSE mechanism. 
While the in-plane $5d_{x^2-y^2}$ orbitals, see Fig.~\ref{Fig1}(b), have $\sigma$-overlap with the bridging oxygen $2p$ orbitals and result in an AF coupling, the out-of-plane degenerate $5d_{xz}$ and $5d_{yz}$ orbitals with $\pi$-type overlap contribute to ferromagnetic exchange that is governed by the Hund's rule coupling of the W $5d$ orbitals. A competition of these two mechanisms result in an overall small AF exchange.  
On the other hand, constraining the virtual hopping into a single bridging W$ ^{6+} $ ion's (with 180$^{\circ}$ Cu-W-Cu angle) unoccupied orbitals result in more than ten times smaller, 0.7 meV, $J_2$ coupling. 
This indicates the predominance of the W virtual orbitals in capturing the $J_2$ coupling.  
Note that there is only one out-of-plane $5d_{xz/yz}$ orbital participating in the hopping via $ \pi $-overlap which along with $\sigma$-type hopping through $5d_{x^2-y^2}$ orbital results in an AF coupling.

\begin{figure}[b!]
	\includegraphics[width=0.98\columnwidth]{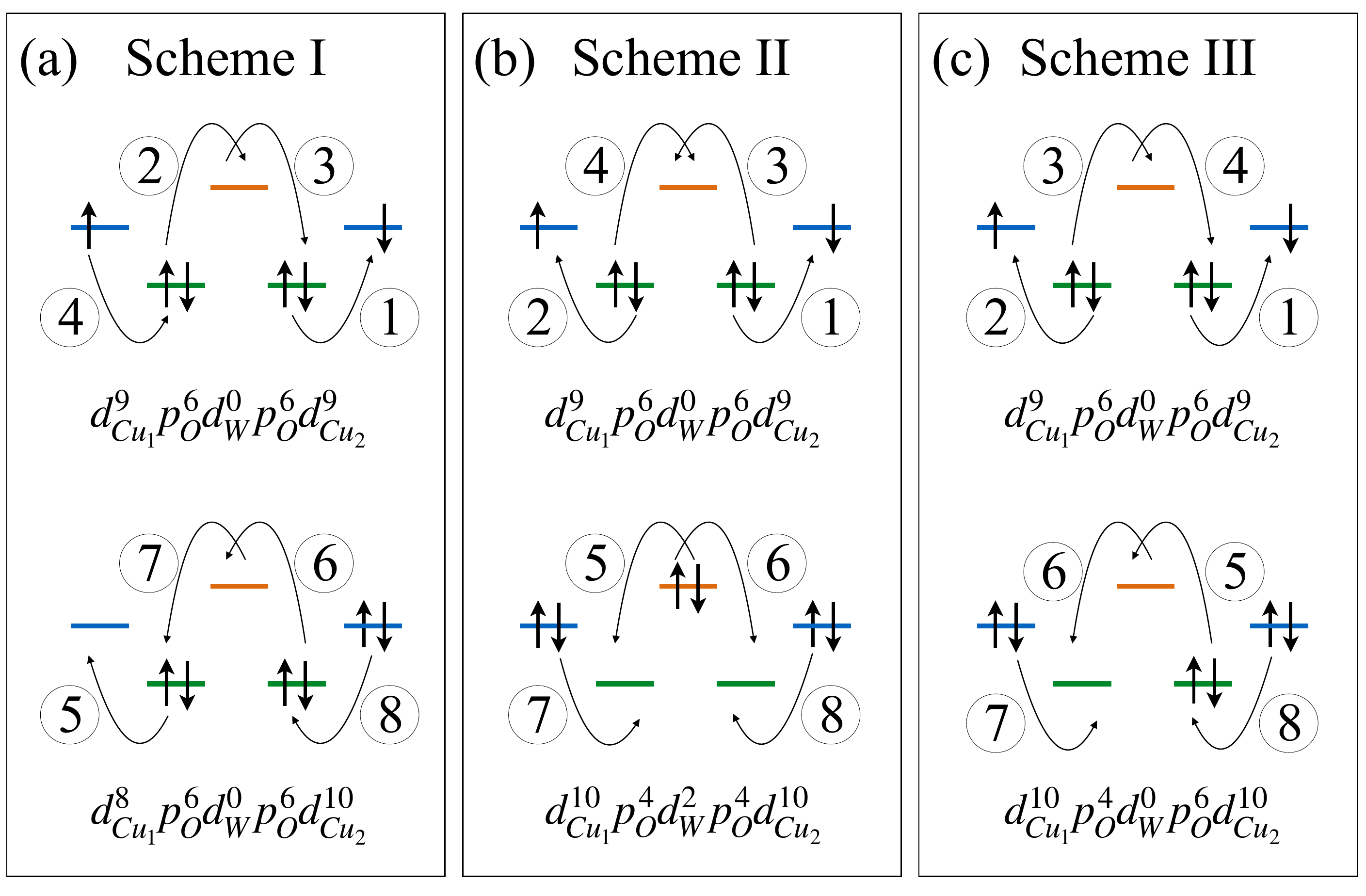}
	\caption{Super-superexchange mechanisms involved in NNN $J_2$ coupling in a four band Hubbard model. The three possible intermediate states (see text) that contribute to exchange interaction are shown as three schemes. The Cu $3d_{x^2-y^2}$, O $2p$ and W $5d_{x^2-y^2}$ levels are represented in blue, green and orange, respectively.  
		The sequence of virtual electron hoppings (arrows) are marked by numbers 1 to 8.  
		%(a) a single hole and a single electron respectively in the O $2p$ and W $5d$ orbitals, (b) both the Os $2p$ and W $5d$  containing two holes and two electrons respectively and (c) only one of the O's $2p$ contains two holes. There are eight virtual hoppings (shown with arrows and labeled with numbers) involved  in each of the mechanisms.}
	}
	\label{Fig_sse}
\end{figure}
We emphasize that in QC calculations all the virtual orbitals of the W$ ^{6+} $ ions participate in the SSE process and estimating the contributions from a particular virtual orbital is impractical. 
 %{\cbl 
 	However, one can understand the SSE from a simplified Hubbard model (SSE-H) that contains two oxygen $p$ orbitals and an additional single W $5d_{x^2-y^2}$ virtual orbital ($d-p-d-p-d$) compared to a conventional $d-p-d$ model applied for charge-transfer insulators~\cite{khomskii-book}.
In Fig.~\ref{Fig_sse}, the SSE processes in the $J_2$ coupling within the SSE-H model are shown. 
There are three different possible virtual hoppings, represented schematically in Fig.~\ref{Fig_sse}, that lift the spin degeneracy and hence contribute to the AFM exchange coupling.  
In scheme \RNum{1}, Fig.~\ref{Fig_sse}(a), the electron from one Cu$ ^{2+} $ ion can hop to the other and back through the intermediate configurations with a single hole (electron) in O $2p$ (W $5d$) orbitals at a particular instance.
This scheme has a dominant contribution to the $J_2$ coupling.
Two other viable possibilities are shown in schemes \RNum{2} and \RNum{3} in  Fig.~\ref{Fig_sse}(b) and ~\ref{Fig_sse}(c). 
Here, configurations where both Cu $ d_{x^2-y^2} $ orbitals are doubly occupied are active. 
While in scheme \RNum{2} both oxygen atoms can contain two holes and the W $5d_{x^2-y^2}$ orbital holds two electrons, in scheme \RNum{3} only one of the oxygen atoms contains two holes. 
The last scheme contributes twice, as either of the two oxygen atoms can accommodate two holes. 
%}

\begin{figure}[t!]
	\includegraphics[width=0.90\columnwidth]{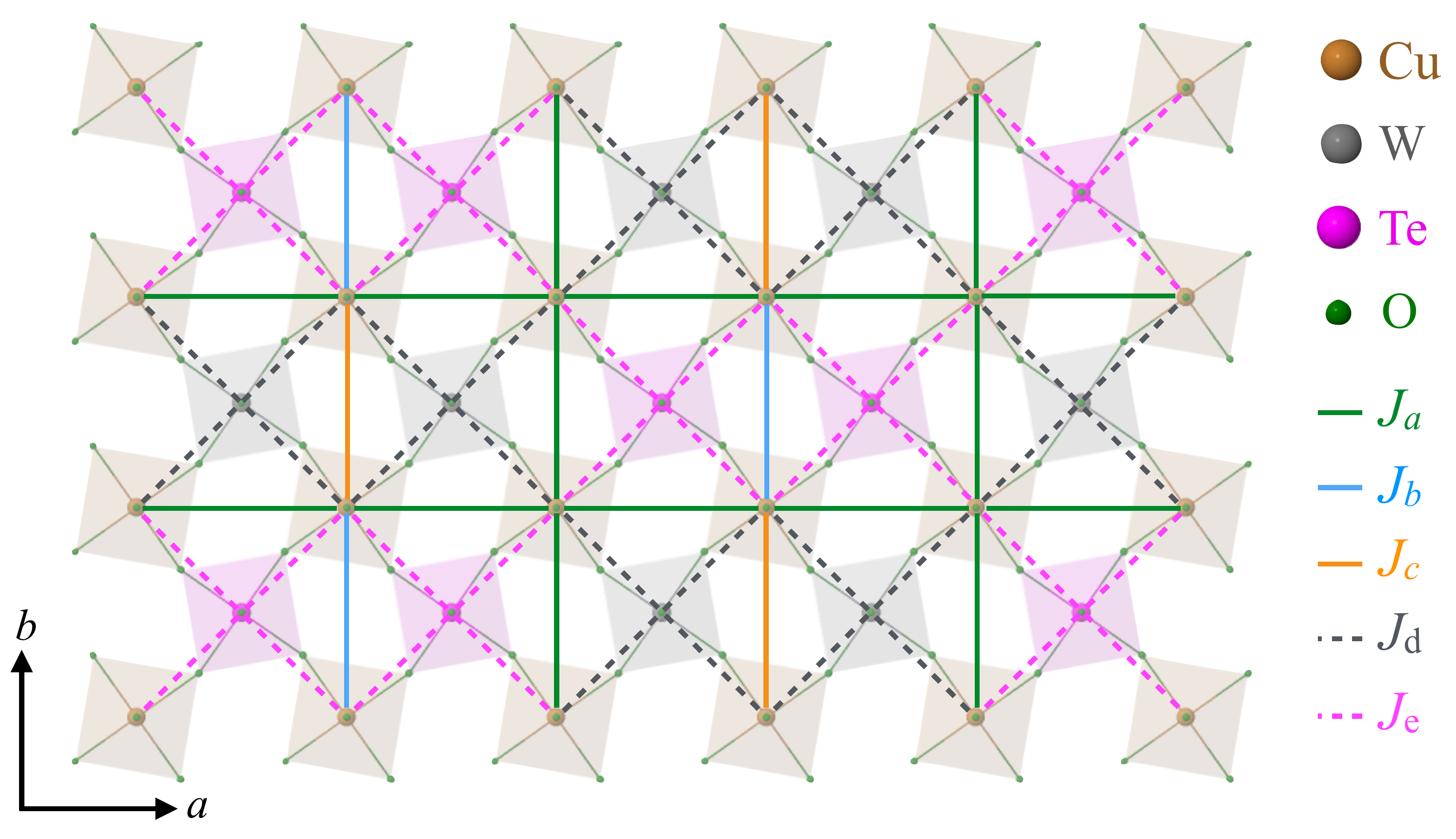}
	\caption{Square lattice scheme with bond disorder and different possible exchange interactions in \SCTWOf.}
	\label{solid_soln}
\end{figure}

The coupling arising from scheme \RNum{1} can be written as
\begin{equation}
J^{\rm \RNum{1}}_2 = 2\frac{t_{dd}^2}{U^{\rm Cu}_{dd}}, \ \mathrm {with} \ \ t_{dd}= \frac{t_{pd_{\rm Cu}}^2}{\Delta_{pd_{\rm Cu}}} \frac{t_{pd_{\rm W}}^4}{\Delta_{pd_{\rm W}}^2},
\end{equation}
where $t_{pd_{\rm Cu}}$ and $t_{pd_{\rm W}}$ are the hopping matrix elements from O $2p$ to Cu $3d_{x^2-y^2}$ and from O $2p$ to W $5d_{x^2-y^2}$, respectively, $U^{\rm Cu}_{dd}$ is the on-site Coulomb interaction on the Cu sites, and $\Delta_{pd_{\rm Cu}}$ and $ \Delta_{pd_{\rm W}} $ are the charge-transfer energies from O $2p$ to Cu $3d$ and W $5d$ orbitals, respectively. 
Schemes \RNum{2} and \RNum{3} would involve $U^{\rm O}_{pp}$ ($U^{\rm W}_{dd}$), the Coulomb interactions when two holes (electrons) are accommodated in O $2p$ (W $5d$) orbitals, and yield minor contributions. 

{\it Effect of Te/W atom disorder on the exchange coupling constants}: 
Let us consider \SCTWOf\ and assume that Te and W atoms are perfectly ordered such that every Cu$^{2+}$ ion is surrounded by two Te$^{6+}$ and two W$^{6+}$ ions. In such a scenario, there are three NN ($J_a$, $J_b$ and $J_c$) and two NNN ($J_d$ and $J_e$) exchange couplings as show in Fig.~\ref{solid_soln}.
Four of the five couplings, $J_b$, $J_c$, $J_d$ and $J_e$, remain the same as in the end compounds \SCWO\ and \SCTO\ as the exchange paths are the same. 
On the other hand, the exchange channels corresponding to $J_a$ are different compared to the end compounds.
    We estimated the coupling $J_a$ from our {\it ab initio} singlet-triplet energy separation for two Cu$^{2+}$ ions with the neighboring environment as shown in Fig.~\ref{solid_soln}. 
    We find this coupling AMF with a  magnitude, 0.3 meV, much smaller than the dominant coupling in the end compounds.
To summarize, our calculations show that the average 
%$\langle J_1 \rangle$ and $\langle J_2 \rangle$ 
$ J_{1} $ and $ J_{2} $
can be tuned from effectively 0 to 8\,meV through substitution of Te for W, opening up an interesting arena to explore bond disorder of a spin-1/2 square lattice antiferromagnet.

% Experimental details

\begin{figure}[b!]
	\centering
	\includegraphics[width=0.99\columnwidth,bb= 14 108 620 560,clip=]{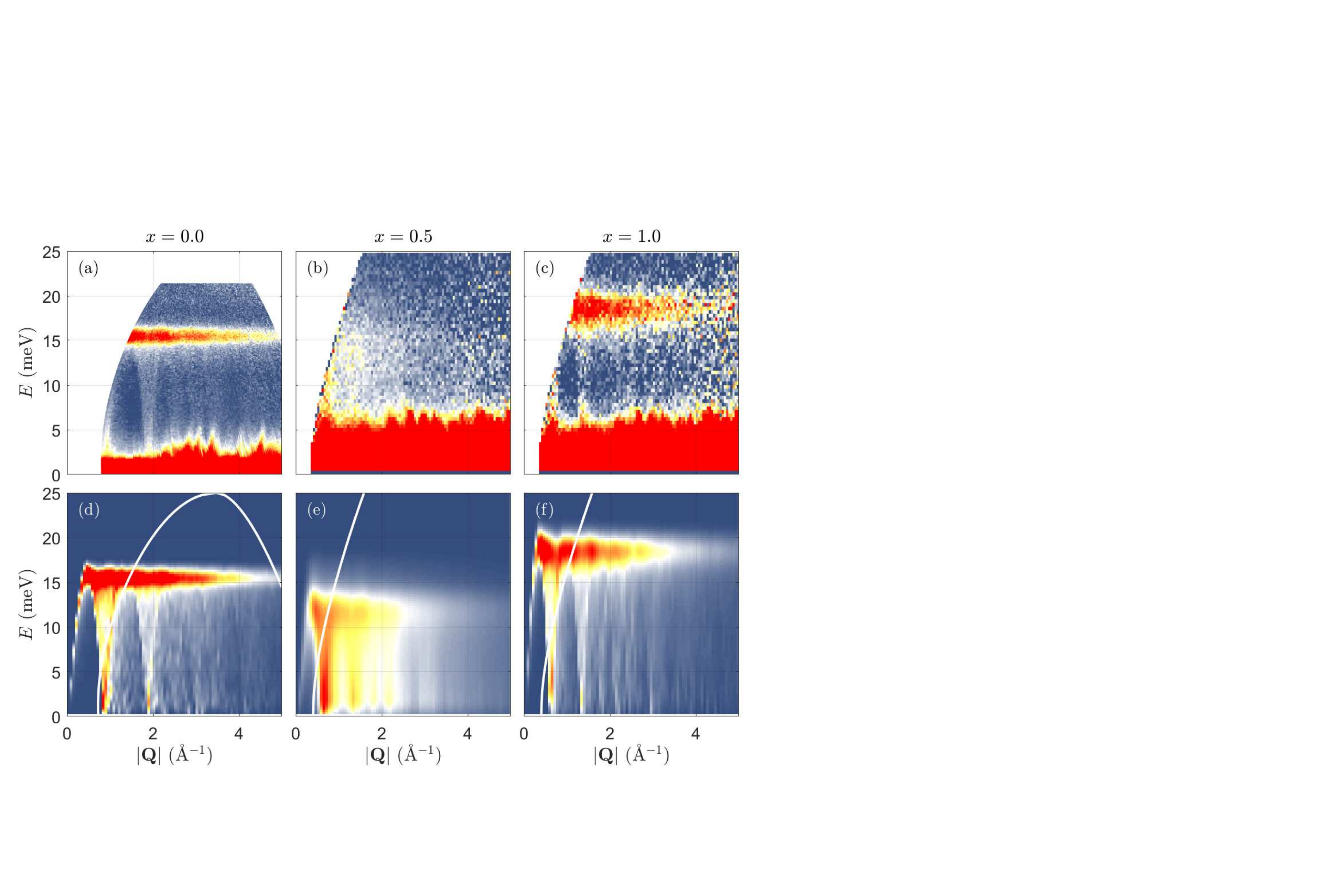}
	\caption{(a)-(c) Measured inelastic powder spectrum $\chi''(\Qb,E)$ of \SCTWO\ for $x = \{0.0,0.5,1.0\}$ compositions. 
		The spectra were normalized by nuclear Bragg scattering for easier comparison. 
		In panels (d)--(f) we present the simulations based on a random exchange-bond order. 
		An energy broadening approximating the instrumental resolution has been applied to each calculated spectra. 
		Solid curves in the simulations represent the range of the detector coverage in the corresponding experiment.}
	\label{Fig3}
\end{figure}

% Spin-wave details
{\it INS experiments and SWT calculations}:
Measurements on \SCTO\ were performed using the IN4 spectrometer at the ILL utilizing an incident neutron energy of $E_i = 25.2$\,meV \cite{cicognani-in4c}. The \SCTWO\, for $x=0.5$ and 1.0 samples were studied using MERLIN at ISIS with $E_i = 45$\,meV \cite{merlin-spectrometer}.  
Further details on \SCTO\ and \SCWO\ INS measurements are reported elsewhere \cite{babkevich_prl_2016, walker-prb-2016}.

	Figure~\ref{Fig3} shows the inelastic neutron scattering spectra of powder \SCTWO\ that have been collected on $x = \{0,0.5,1.0\}$ compounds. 
	A background, adjusted for the Bose thermal population factor, recorded at $>$100 K has been subtracted from the spectra to remove the phonon contribution at larger \Qb. 
	The end-compounds of \SCWO\ and \SCTO\ show spin waves dispersing from the CAF and NAF zone centres, respectively. 
	A strong band of scattering around 15-17 meV is found in both compounds. 
	This corresponds to a van Hove singularity from the top of the spin-wave dispersion. 
	The INS spectrum of the intermediate \SCTWOf\ compounds is dramatically different. 
	There appears to be a significant smearing of the spectrum in momentum and energy transfer. 
	The band of scattering corresponding to the van Hove singularity is absent. 
	Weak excitations are observed up to around 20\,meV. 
	This scattering decreases with increasing \Qb, as would be expected for magnetic scattering. 
	Magnetic modes emerge from $\Qb = 0.65$ and 1.4\,\AA$ ^{-1} $, much like in \SCWO, which would suggest the dominant interactions in \SCWO\ persist in the $x=0.5$ compound.

To simulate the INS spectra, we need to construct an appropriate magnetic ground state from which magnetic fluctuations can be calculated.
We define a $10\times10$ square lattice with randomly populated W and Te atoms. 
The strengths of the $J_1-J_2$ exchange parameters are as given in Table~\ref{Exchg_inter} and the different possible exchange pathways in the mixed $x=0.5$ compound are according to Fig.~\ref{solid_soln}. 
Therefore, $J_1$ can take values of 7.60\,meV or 1.02\,meV depending on whether two Te or W atoms are involved in the exchange process with similar arguments applying to $J_2$. 
In the case of one W and one Te atom, we take $J_1 = 0$. 
From this construction, we find the classical spin configuration which minimises the total energy and calculate the spin-wave dispersion. 
To account for truncation of the spin Hamiltonian at the quadratic terms when calculating the one-magnon energy we rescale the magnon energy by a constant factor of $Z_c =1.18$~\cite{singh-prb-1989}.
The calculation is repeated with different distributions of Te and W and the resulting spin-wave pattern is averaged.
Figures~\ref{Fig3}(d)-\ref{Fig3}(f) show the calculated powder averaged spectra for each composition. 
Comparing the calculated spectra for $x=0.5$ to the end compounds, we observe that the simulation predicts a rather broad spectrum. 
The intense and sharp scattering at the top of the bandwidth in \SCWO\ and \SCTO\ is no longer present for the intermediate compound. 
Therefore, the effect of the substitution for the intermediate compounds is to leave the powder spectrum featureless with the exception of excitations that emerge from Neel-like and CAF-like low-energy excitations centered at $\Qb \approx 0.7$ and 1.4\,\AA$ ^{-1} $.
The spectrum, despite the lack of long range magnetic order and contrary to expectations of the limitations of LSWT~\cite{Zhang_prl_2019}, appears to be in good agreement with the measured powder spectrum, indicating that the bond-disordered exchange couplings reproduce the INS spectrum of \SCTWOf.

To summarize, we have computed the NN and NNN Heisenberg exchange couplings in \SCWO\ and  \SCTWOf\ finding excellent agreement with available experimental observations. 
We established that the non-magnetic cation bridging the magnetic sites play a significant role in the SSE process. 
In the case of a completely filled $d$-manifold (Te$ ^{6+} $) cation, the exchange path does not include any of its orbitals, but for the $d^0$ (W$^{6+}$) bridging cation, the SSE process via these empty orbitals is pivotal. 
While these conclusions corroborate with DFT+$U$ based studies~\cite{Yuanhui_DFT_2017,Vasala_prb_2014, walker-prb-2016}, it is important to know that the computed exchange couplings strongly depend on the choice of the Coulomb repulsion parameter $U$. 
We further provided the rationale for the observed exchange interactions, justifying with numerical evidence. 
Our simulated INS spectra for \SCTO\ and \SCWO\ compare extremely well with experimental data, and they give a good understanding of the measured powder spectrum for \SCTWOf.
Although further neutron scattering studies are necessary to examine the latter compound, our calculations provide a deep insight into the nature of the interactions within the complex ground state of this system.
%{\cbl Investigation of other magnetic systems with exchange over non-magnetic cations, e.g. Cr$_2$Mo/TeO$_6$ ~\cite{Zhu_cr_w_te_prl_2014} 
%	and Ba$_2$Y/InOsO$_6$~\cite{Feng_os_dp_2019} will enable us to formulate the general rules for SSE phenomenon. }
Our work thus establishes the theoretical background for describing bond-disorder exchange couplings highlighting site-disordered materials as a new playground for exploring QSL states.

\begin{acknowledgments}
    V.M.K. and O.V.Y. acknowledge the support from ERC project `TopoMat' (grant No. 306504), Swiss NSF NCCR MARVEL and SNSF Sinergia grant CRSII5\_171003. 
    The authors would like to acknowledge CSCS (project s832) and EPFL-SCITAS for providing the computational resources. 
\end{acknowledgments}

%\bibliography{shorttitles,biblio_V8}

\end{document}